\documentclass[12pt]{JHEP3}

\usepackage{amscd,amsmath,amssymb,amsfonts,xspace,mathrsfs}
\usepackage{graphicx}
\usepackage{fancyhdr}
\usepackage{rotating}
\usepackage{psfrag}
\usepackage{epsfig}
\usepackage{url}
\usepackage{bbold}

\allowdisplaybreaks
\newcounter{tabl}
\newcommand{\be}{\begin{equation}}
\newcommand{\ee}{\end{equation}}
\def\ba{\begin{eqnarray}}
\def\ea{\end{eqnarray}}
\newcommand{\beq}{\begin{eqnarray}}
\newcommand{\eeq}{\end{eqnarray}}
\newcommand{\bea}[2]{\be\label{#2}\begin{array}{#1}}
\newcommand{\eea}{\end{array}\ee}

\def\diag{{\rm diag}}

\def\({\left(}
\def\){\right)}
\def\[{\left[}
\def\]{\right]}
\def\p{\partial}
\newcommand{\de}{\mathrm{d}}

\def\11{1\!\! 1}

\def\hf{\frac{1}{2}}

\def\eps{\varepsilon}

   \def\CC {{\cal C}}

   \def\CG {{\cal G}}
   \def\CH {{\cal H}}

   \def\CN {{\cal N}}

   \def\CQ {{\cal Q}}
   
   \def\CS {{\cal S}}

   \def\CW {{\cal W}}
   \def\CX {{\cal X}}

\newcommand{\opB}{\mathfrak{B}}
\newcommand{\opD}{\mathfrak{D}}
\newcommand{\frU}{\mathfrak{U}}
\newcommand{\frV}{\mathfrak{V}}

\newcommand{\tX}{\lefteqn{\smash{\mathop{\vphantom{<}}\limits^{\;\sim}}}X}
\newcommand{\tCX}{\lefteqn{\smash{\mathop{\vphantom{<}}\limits^{\;\sim}}}\CX}

\newcommand{\tP}{\lefteqn{\smash{\mathop{\vphantom{<}}\limits^{\;\sim}}}P}

\newcommand{\tp}{\lefteqn{\smash{\mathop{\vphantom{\scriptstyle{c}}}\limits^{\sim}}}p}

\def\hCG{\hat\CG}

\def\hN{\hat N}

\def\xc{{\rm x}}
\def\yc{{\rm y}}

\def\omp{\omega_+}
\def\omm{\omega_-}
\def\ompm{\omega_\pm}

\def\ompi#1{\omega_{+,#1}}
\def\ommi#1{\omega_{-,#1}}
\def\ompmi#1{\omega_{\pm,#1}}

\def\hom{\hat\omega}

\def\tXp{\tX_+}
\def\tXm{\tX_-}
\def\tXpm{\tX_\pm}

\def\tXpi#1{\tX_{+,#1}}
\def\tXmi#1{\tX_{-,#1}}
\def\tXpmi#1{\tX_{\pm,#1}}

\def\tCXp{\tCX_+}
\def\tCXm{\tCX_-}
\def\tCXpm{\tCX_\pm}

\def\eep{e_+}
\def\eem{e_-}
\def\eepm{e_\pm}

\def\eepi#1{e_{+,#1}}
\def\eemi#1{e_{-,#1}}
\def\eepmi#1{e_{\pm,#1}}
\def\eempi#1{e_{\mp,#1}}

\def\tpp{\tp_+}
\def\tpm{\tp_-}
\def\tppm{\tp_\pm}

\def\tppi#1{\tp_{+,#1}}
\def\tpmi#1{\tp_{-,#1}}
\def\tppmi#1{\tp_{\pm,#1}}
\def\tpmpi#1{\tp_{\mp,#1}}

\def\Np{N_+}
\def\Nm{N_-}
\def\Npm{N_\pm}

\def\gp{g_+}
\def\gm{g_-}
\def\gpm{g_\pm}

\def\gpi#1{g_{+,#1}}
\def\gmi#1{g_{-,#1}}

\def\cfQpm{\CQ_{\pm}}

\def\cfQpi#1#2{\CQ_{+,#1}^{\,#2}}
\def\cfQmi#1#2{\CQ_{-,#1}^{\,#2}}
\def\cfQpmi#1#2{\CQ_{\pm,#1}^{\,#2}}

\def\mett{{\mathbf g}}
\def\metb{{\mathbf h}}

\def\Dp{D^{(+)}}
\def\Dm{D^{(-)}}
\def\Dpm{D^{(\pm)}}

\def\PhiG{{\cal G}}
\def\PhiC{{\cal C}}

\def\CGp{\PhiG_+}
\def\CGm{\PhiG_-}
\def\CGpm{\PhiG_\pm}

\def\CGpi#1{\PhiG_{+,#1}}
\def\CGmi#1{\PhiG_{-,#1}}

\def\CCp{\PhiC_+}
\def\CCm{\PhiC_-}
\def\CCpm{\PhiC_\pm}

\def\CCpi#1{\PhiC_{+,#1}}
\def\CCmi#1{\PhiC_{-,#1}}

\def\hCC{\hat\PhiC}

\def\Ham{\CH_{\rm tot}}
\def\Hami{H_{\rm tot}}

\def\Ns{N_{\rm n.d.}}
\def\dth{{\varrho}}

\title{On Partially Massless Theory in 3 Dimensions}

\author{Sergei Alexandrov$^{1,2}$ and C\'edric Deffayet$^{3,4}$
\\
$^1$ {\it Universit\'e Montpellier 2, Laboratoire Charles Coulomb UMR 5221, F-34095,
Montpellier, France}\\
$^2$ {\it CNRS, Laboratoire Charles Coulomb UMR 5221, F-34095,
Montpellier, France}\\
$^3$ {\it UPMC-CNRS, UMR7095, Institut d'Astrophysique de Paris,
GReCO, 98bis boulevard Arago, F-75014 Paris, France}
\\
$^4$ {\it IH\'ES, Le Bois-Marie, 35 route de Chartres, F-91440 Bures-sur-Yvette, France}

\vspace*{2mm} {\tt e-mail:
\email{salexand@univ-montp2.fr},
\email{deffayet@iap.fr}
}

}

\abstract{We analyze the first-order formulation of the ghost-free bigravity model in three-dimensions known as zwei-dreibein gravity.
For a special choice of parameters, it was argued to have an additional gauge symmetry and give rise to a partially massless theory.
We provide a thorough canonical analysis and identify that whether the theory becomes partially massless depends on
the form of the stability condition of the secondary constraint responsible for the absence of the ghost.
Generically, it is found to be an equation for a Lagrange multiplier
implying that partially massless zwei-dreibein gravity does not exist.
 However, for special backgrounds this condition is identically satisfied leading to the presence of
additional symmetries, which however disappear at quadratic order in perturbations.}

\begin{document}

\section{Introduction}

Massive gravity (see \cite{Rubakov:2008nh,Hinterbichler:2011tt,deRham:2014zqa} for reviews) has recently attracted
a lot of attention with motivations ranging from cosmology
\cite{Dvali:2000hr,Deffayet:2001pu,Deffayet:2000uy,Gabadadze:2003jq,D'Amico:2011jj}
to more abstract ones \cite{Bergshoeff:2009hq}.  Since the seminal work of Fierz and Pauli \cite{Pauli:1939xp,Fierz:1939ix},
it is well known that a sound massive graviton on a four dimensional flat space-time has five propagating
positive energy polarizations. Going beyond the linear Fierz-Pauli theory while keeping the same number
of polarizations has long been thought impossible \cite{Boulware:1973my,Creminelli:2005qk}.
However, it was recently discovered that such a non-linear extension does exist,
which is now known as dRGT theory \cite{deRham:2010ik,deRham:2010kj}.
The absence of a problematic sixth ghostlike mode, known as the Boulware-Deser (BD) ghost,
has first been shown in the so-called decoupling limit
\cite{deRham:2010ik,deRham:2010kj,deRham:2011rn} (using in particular the observations of
\cite{Creminelli:2005qk,Deffayet:2005ys}) and then in the full non-linear setup \cite{Hassan:2011hr,Hassan:2011ea}.

In its original form, the dRGT theory is expressed via two metrics, one dynamical and the other non-dynamical and flat.
However, it can easily be extended to a bimetric theory where
both metrics are dynamical as well as to an arbitrary number of dimensions \cite{Hassan:2011ea,Hassan:2011zd}.
The resulting ghost-free bimetric gravity
has degrees of freedom corresponding to one massless graviton and one massive one devoid of the BD ghost.
E.g. in four dimensions, this corresponds to a total of 2+5 degrees of freedom,
while in three dimensions it just gives 2 degrees of freedom
(i.e. counting 0 for a massless graviton and 2 for a massive graviton).
It is the three-dimensional version of this bigravity theory that will be our main interest here.

Before we turn to this concrete model, let us further discuss some features
of massive gravitons on non-flat backgrounds. It is known for a long time that
the linear theory of a massive gravity can differ drastically from linearized
General Relativity depending on the background metric.
For instance, on the flat background one does not recover predictions of the massless theory (i.e. linearized General Relativity)
in the massless limit --- this fact is the celebrated van Dam-Veltman-Zakharov (vDVZ)
discontinuity \cite{vanDam:1970vg,Zakharov:1970cc,Iwasaki:1971uz},
which is believed to be removed by taking into account non-linear effects,
thanks to the so-called "Vainshtein mechanism" \cite{Vainshtein:1972sx,Deffayet:2001uk,Babichev:2009us,Babichev:2009jt,Babichev:2010jd}
(see e.g. \cite{Babichev:2013usa} for an introduction).
On the other hand, the vDVZ discontinuity is absent whenever the background metric is taken to be dS or AdS
\cite{Higuchi:1986py,Porrati:2000cp,Kogan:2000uy}.
However, on de Sitter spacetime a new phenomenon appears --- the sign of
the kinetic term of one of the graviton modes flips whenever
the mass of the graviton $m$ crosses the Higuchi bound $m_{\rm H}^2 = 2\Lambda/3 $ \cite{Higuchi:1986py,Higuchi:1989gz}
where $\Lambda$ is the cosmological constant. In the case where the mass saturates the bound, $m=m_{\rm H}$,
the gravitons (in four dimensions) turn out to have only 4 polarizations, i.e. one polarization less than away from this point.
Such a theory is known now as partially massless gravity
\cite{Deser:1983mm,Deser:2001pe,Deser:2001us,Deser:2001wx,Deser:2001xr,Deser:2004ji}.

The disappearance of one polarization in the partially massless case happens due to an extra symmetry which arises at the Higuchi point.
A fascinating question is whether this symmetry, and hence the absence of one of the graviton modes,
can be extended to a fully non-linear theory of massive gravity.
This problem is of great importance because massive gravity, even in the absence of the BD ghost,
has various pathological properties. To name a few, it is known
that it has a hidden very low strong interaction scale which makes its UV completion problematic;
there are strong arguments that the dRGT model suffers from superluminality in
the decoupling limit \cite{Gruzinov:2011sq,deFromont:2013iwa,Deser:2013eua};
and last but not least, its cosmological solutions have been shown to be generically unstable \cite{DeFelice:2012mx,DeFelice:2013bxa}.
It was argued that partial masslessness can provide a healthy resolution, or at least an improvement,
of most of these pathologies \cite{Deser:2013gpa}.
Therefore, a positive answer to the above question might have important physical implications.

An obvious playground to investigate this issue is the dRGT theory or its bimetric version suggested in \cite{Hassan:2011zd}, 
given that they bear the mode counting of the Fierz-Pauli setup to a full non-linear one.
This problem has already been considered by several authors, reaching sometimes different conclusions \cite{deRham:2012kf,Fasiello:2012rw,Hassan:2012gz,Joung:2012hz,Hassan:2012rq,Deser:2013xb,Deser:2013bs,Deser:2013uy,deRham:2013wv,Joung:2014aba,Hassan:2013pca}.
It should be noted that, whereas obstructions to partial masslessness
and the pathological properties mentioned in the previous paragraph
have been mostly discussed in the framework of massive gravity, the works \cite{Hassan:2012gz,Hassan:2012rq,Hassan:2013pca}
and, in particular, \cite{Hassan:2014vja} pointed out that there might be substantial differences in the dynamics of
massive gravity and bigravity models.
Furthermore, the authors of \cite{Hassan:2012gz,Hassan:2012rq} singled out a subset of
ghost-free bimetric models, in which all parameters of the interaction potential are determined by the value
of the cosmological constant, as the only sensible candidates for non-linear partially massless theories.
It is important to emphasize that the criterions obtained in \cite{Hassan:2012gz,Hassan:2012rq} can only be satisfied
if both metrics are taken to be dynamical and if the number of spacetime dimensions is 3 or 4.\footnote{In \cite{Hassan:2012rq}
it was shown that in higher dimensions partial masslessness, if it exists, requires inclusion of higher curvature terms.}

The purpose of this work is to provide a definite answer to the question of the existence of
a non-linear partially massless theory in the simplest case of three dimensions.
To this end, we perform a thorough canonical analysis of the three-dimensional ghost-free bigravity model
in the first-order formalism. A reformulation of ghost-free massive gravity and bigravity models using vielbeins
has been suggested in \cite{Hinterbichler:2012cn}. It makes these models much simpler and more elegant
replacing the complicated dRGT potential involving a matrix square root by a polynomial one.
Although there are some subtle issues concerning the full equivalence of this
vielbein formulation with the original form of the theory in terms of the metric \cite{Deffayet:2012zc,Banados:2013fda},
for generic values of the parameters of the interaction potential the complete canonical analysis
of the first order four-dimensional bimetric gravity
has been performed in \cite{Alexandrov:2013rxa} and confirmed that the theory is ghost free
(see also \cite{Deffayet:2012nr,Alexandrov:2012yv,Comelli:2013txa,Kluson:2013aca,Deser:2014hga}).

The three-dimensional version of the general construction of \cite{Hinterbichler:2012cn} has been analyzed in \cite{Bergshoeff:2013xma}
under the name of zwei-dreibein gravity, but the issue of partial masslessness was not investigated.
Here we go beyond the previous analysis by computing the Dirac algebra of constraints in full detail.
This allows us to identify the condition which governs whether the theory is partially massless or not.
This turns out to depend on the role played in the canonical formulation by
the stability condition of the secondary constraint, which is one of the two constraints removing
the Boulware-Deser ghost.
The theory becomes partially massless only if this condition is either automatically satisfied
or leads to a tertiary constraint.
However, we show that neither of these two possibilities is realized and
thus the three-dimensional ghost-free bigravity does not feature partial masslessness.

We emphasize that this result is obtained at full non-linear level. On the other hand, a linearization
around a background spacetime (in fact, {\it two} spacetimes since our model contains {\it two} dynamical metrics)
can drastically change the canonical structure and thereby lead to the appearance
of partial masslessness. We show that it does happen for a class of backgrounds which includes, in particular,
the so called ``proportional backgrounds".
These results provide an explanation and a generalization of the observations made in \cite{Hassan:2012gz,Hassan:2012rq}.

Our work is organized as follows. In the next section, we introduce the model,
its Lagrangian and Hamiltonian formulations. In particular, we identify all constraints and their nature.
In section \ref{sec-pm} we observe how partial masslessness could potentially appear,
but after a careful analysis we prove that in fact it does not occur. Then in section \ref{sec-backgr}
we show that for linearization around specific backgrounds the original symmetries of the model are enhanced,
which however is just an artefact of the linearization.
Section \ref{sec-concl} is devoted to conclusions. Three appendices contain our conventions and technical details.

\section{Zwei-Dreibein gravity and its canonical analysis}

\subsection{Lagrangian formulation}

The zwei-dreibein gravity model \cite{Hinterbichler:2012cn,Bergshoeff:2013xma} is given by the following action
(see appendix \ref{ap-conv} for our conventions)
\be
S[\eepm,\ompm]=S_{\rm HP}[\eep,\omp]+S_{\rm HP}[\eem,\omm]+S_{\rm int}[\eep,\eem],
\label{totalaction}
\ee
where the dynamics of each of the two sectors is described by the usual three-dimensional Hilbert-Palatini action
of the first-order formalism\footnote{For simplicity, we absorbed the two Newton constants by rescaling
the tetrad fields $\eepm^I$ and the cosmological constants $\Lambda_\pm$ in the two sectors.
They can always be restored by dimensional analysis.}
\be
S_{\rm HP}[e,\omega]=\int \(e^I\wedge  F_I(\omega)-\frac{\Lambda}{6}\,\eps_{IJK} e^I\wedge e^J\wedge e^K\),
\label{HPaction}
\ee
whereas the coupling between them is given by the interaction term
\be
S_{\rm int}
=-\hf\int \eps_{IJK}\(\beta_+\eep^I\wedge \eep^J\wedge \eem^K
+\beta_-\eep^I\wedge \eem^J\wedge \eem^K\).
\label{intpot}
\ee
If at least one of the two coupling constants $\beta_\pm$ is non-vanishing, the theory was shown to describe
ghost-free bimetric gravity propagating one massless and one massive gravitons.\footnote{In \cite{Banados:2013fda}
it was noticed that the absence of the ghost holds only under assumption of a certain non-degeneracy condition,
not following from invertibility of the two tetrads $e_\pm$. We will meet this condition in the next subsection.}

The main question, which we are interested in, however is whether the theory \eqref{totalaction} becomes partially massless
for a suitable choice of parameters analogous to the saturation of the Higuchi bound \cite{Higuchi:1986py,Higuchi:1989gz}.
In \cite{Hassan:2012gz,Hassan:2012rq} a one-parameter candidate for partially massless theory was identified in the metric formalism
by analyzing linear perturbations around the so-called proportional backgrounds
\be
\gmi{\mu\nu}=c^2\gpi{\mu\nu}.
\label{propbackmet}
\ee
It corresponds to the case where the coupling and cosmological constants are all equal
\be
\Lambda_+=\beta_+=\beta_-=\Lambda_-\equiv\Lambda.
\label{pmchoice}
\ee
For this special choice of parameters the action \eqref{totalaction} indeed crucially simplifies.
Introducing diagonal and off-diagonal fields  $e$, $\hat{e}$, $\omega$ and $\hat{\omega}$ according to
\be
\eepm^I=\hf\(e^I\pm \hat e^I\),
\qquad
\ompm^I=\omega^I\pm \hom^I,
\label{defdiagfields}
\ee
this action, whenever (\ref{pmchoice}) is satisfied, can be rewritten as
\be
S_{\rm p.m.}[e,\omega,\hat e, \hom]=S_{\rm HP}[e,\omega]+\int\(\hat e^I \wedge D\hom_I+\hf\, \eps_{IJK} e^I\wedge \hom^J\wedge \hom^K\).
\ee
Thus, the diagonal variables $e^I$ and $\omega^I$ describe the usual three-dimensional gravity, whereas $\hat e^I,\hom^I$ correspond
to a massive field with only one non-standard feature that its mass is affected by the metric through a non-minimal coupling to $e^I$.

\subsection{Canonical formulation}

The canonical analysis of the action \eqref{totalaction} has been already performed in
\cite{Bergshoeff:2013xma,Banados:2013fda,Bergshoeff:2014bia}. However, the constraint algebra was not computed in full details,
and the possibility of various degeneracies
 of the Dirac matrix, and as a consequence, a change of the canonical structure
for special choices of parameters was not investigated. This is precisely our goal in this paper.
Therefore, we repeat the main steps of the canonical analysis, meanwhile setting up our notations
and deriving the results necessary for the further analysis of partial masslessness
which will be given in section \ref{sec-pm}.
Accordingly, we first do not assume that (\ref{pmchoice}) holds. Our presentation mainly follows \cite{Alexandrov:2013rxa}
where a similar canonical analysis was done for the four-dimensional model.

Splitting all spacetime indices into time and spatial components, the action \eqref{totalaction}
can be put into the canonical form
\be
S=\int \de t\de^2\xc\(\tppi{I}^a\p_t\ompi{a}^I+\tpmi{I}^a\p_t\ommi{a}^I-\Ham\),
\ee
where we defined the variables
\be
\tppmi{I}^a\equiv \eta_{IJ}\eps^{ab}\eepmi{b}^J
\ee
playing the role of the momenta canonically conjugated to $\ompmi{a}^I$, respectively.
The Hamiltonian
\be
\Ham =-\ompi{0}^I\CGpi{I}-\ommi{0}^I\CGmi{I}-\eepi{0}^I\CCpi{I}-\eemi{0}^I\CCmi{I}
\label{hamilt}
\ee
is given by a linear combination of the primary constraints
\beq
\CGpi{I}&=& \Dp_a\tppi{I}^a,
\\
\CGmi{I}&=& \Dm_a\tpmi{I}^a,
\\
\CCpi{I}&=&F_I(\omp)-{\eps_I}^{JK}\eps_{ab}\(\frac{\Lambda_+}{2}\,\tppi{J}^a\tppi{K}^b
+\beta_+\tppi{J}^a\tpmi{K}^b+\frac{\beta_-}{2}\,\tpmi{J}^a\tpmi{K}^b\),
\\
\CCmi{I}&=&F_I(\omm)-{\eps_I}^{JK}\eps_{ab}\(\frac{\Lambda_-}{2}\,\tpmi{J}^a\tpmi{K}^b
+\beta_-\tppi{J}^a\tpmi{K}^b+\frac{\beta_+}{2}\,\tppi{J}^a\tppi{K}^b\),
\eeq
where $\Dpm$ denotes the covariant derivative defined by the connection $\ompm$.

Next, we have to study the stability of the primary constraints under time evolution.
To this end, we compute their algebra using the canonical commutation relations
between the two pairs of canonical variables $(\tppmi{I}^a,\ompmi{a}^I)$.
The result is presented in appendix \ref{ap-algebra}.
It immediately implies that the diagonal Gauss constraints $\CG_I=\CGpi{I}+\CGmi{I}$ weakly commute with all primary constraints
and therefore they are automatically stable.

On the other hand, commuting the off-diagonal Gauss constraint $\hCG_I=\CGpi{I}-\CGmi{I}$ with the Hamiltonian
$\Hami=\int\de^2 \xc\,\Ham$,
one finds the following result
\be
\p_t \hCG_I= \{\hCG,\Hami\}
\approx
2\eps_{ab}\Bigl(
\tP^a_I
\((\tpm^b\eepi{0})-(\tpp^b\eemi{0})\)
-E_{0,I}(\tpp^a\tpm^b)\Bigr),
\label{staboffdG}
\ee
where we introduced the weighted diagonal fields
\be
E_\mu^I=\beta_+\eepi{\mu}^I+\beta_-\eemi{\mu}^I,
\qquad
\tP^a_I\equiv \beta_+\tppi{I}^a+\beta_-\tpmi{I}^a.
\label{defEP}
\ee
In the following we will assume that not only the triads $\eepmi{\mu}^I$, but also the diagonal one $E_\mu^I$ is non-degenerate,
i.e. invertible matrix.\footnote{Note that, as shown in \cite{Banados:2013fda}, this assumption
is not guaranteed by the field equations, but it is crucial for the correct counting of d.o.f.,
and hence for the elimination of the BD ghost.}
Then, equating \eqref{staboffdG} to zero, one finds that the three equations can be split into
two equations fixing Lagrange multipliers $\eepmi{0}^I$
\be
(\tpm^b\eepi{0})=(\tpp^b\eemi{0})
\label{cone0}
\ee
and one secondary constraint
\be
\CS=\eps_{ab}(\tpp^a\tpm^b)=0.
\label{defS}
\ee
Altogether these conditions are equivalent to the covariant equations
\be
\eta_{IJ}\eepi{[\mu}^I\eemi{\nu]}^J=0
\ee
which are known as symmetricity  conditions \cite{Hinterbichler:2012cn,Deffayet:2012zc,Banados:2013fda}
ensuring the equivalence of the metric and tetrad formulations of bimetric gravity.

The stability of the remaining two sets of primary constraints generates the following conditions
\be
\begin{split}
\p_t \CCp^{I} \approx &\,
-2\Bigl(\eta^{IJ}\tpmi{J}^a\((E_a\hom_0)-(E_0\hom_a)\)+\eemi{0}^I(\tP^a\hom_a)\Bigr)=0,
\\
\p_t \CCm^{I} \approx &\,
2\Bigl(\eta^{IJ}\tppi{J}^a\((E_a\hom_0)-(E_0\hom_a)\)+\eepi{0}^I(\tP^a\hom_a)\Bigr)=0,
\end{split}
\ee
written in terms of fields \eqref{defEP} and $\hom_\mu^I$ defined in \eqref{defdiagfields}.
Both equations lead to the same two conditions on the Lagrange multipliers $\hom_0^I$
\be
(E_a\hom_0)=(E_0\hom_a),
\label{conEom}
\ee
and to another secondary constraint
\be
\Psi=(\tP^a\hom_a)=0.
\label{defPsi}
\ee

As a result, the stabilization procedure of the primary constraints has led to the two secondary constraints,
$\CS$ \eqref{defS} and $\Psi$ \eqref{defPsi}, which also must be stabilized.
The corresponding conditions can be obtained using \eqref{algsecond}. For $\CS$, it reads
\be
\p_t\CS\approx
-2\({\eps_I}^{JK}\eps_{ab}\tppi{J}^a\tpmi{K}^b\hom_0^I +{\eps^{I}}_{JK}\(\tpmi{I}^a\eepi{0}^J+\tppi{I}^a\eemi{0}^J\)\hom_a^K\)=0,
\label{stabS}
\ee
where we used the Gauss constraints $\CGpm^I$ and the condition \eqref{cone0}. This equation fixes one component of $\hom_0^I$.
Thus, altogether \eqref{conEom} and \eqref{stabS} completely determine these Lagrange multipliers.

At the same time, the stability condition for $\Psi$
\be
\p_t\Psi=\{\hCG(\hom_0)+\CCp(\eepi{0})+\CCm(\eemi{0}),\Psi\}=0
\label{stPsi}
\ee
is expected to generate a condition fixing a certain combination of the Lagrange multipliers $\eepmi{0}^I$
since $\CCpm^I$ are both non-commuting with $\Psi$ (see \eqref{algsecond}).

This completes the (naive) canonical analysis. Since \eqref{conEom} and \eqref{stabS} allow to determine all $\hom_0^I$,
whereas \eqref{cone0} and \eqref{stPsi} fix three components of $\eepmi{0}^I$, the set of first class constraints should
consist of the diagonal Gauss $\CG_I$ and three constraints constructed from $\CCpm^I$. All other constraints including
$\hCG^I$, $\CS$, $\Psi$ and the other three combinations of $\CCpm^I$ should be second class.
This leads to the dimension of the phase space given by $4\times 6-2\times 6-8=4$, which corresponds
to the 2 degrees of freedom of massive graviton in three dimensions \cite{Bergshoeff:2013xma}.

\section{Condition for partial masslessness}
\label{sec-pm}

\subsection{Possible scenarios}
\label{subsec-scenario}

The analysis presented in the previous section relies on the assumption that the stability condition \eqref{stPsi}
can always be solved with respect to one of the Lagrange multipliers. But is it really true?
To check whether this is the case, one should compute this condition explicitly and impose there all the constraints
and equations fixing Lagrange multipliers found at previous stages. This is the goal of this section.

Before we go into details of the calculation, let us outline what are the different possibilities
that can be realized and which canonical structures they would imply.
\begin{itemize}
\item
The simplest case is the one assumed above when the stability condition induces an equation
for one of the Lagrange multipliers.
The corresponding canonical structure is summarized in the end of the previous section and corresponds to
the usual ghost-free bigravity.

\item
The next possibility is obtained if the stability condition takes the form
\be
\CN(\eepmi{0}) \Upsilon=0,
\label{degstabcon}
\ee
where $\CN$ is a linear combination of the Lagrange multipliers $\eepmi{0}^I$ and $\Upsilon$ is a function of canonical variables.
If one of the non-degeneracy conditions imposed on the theory implies that $\CN$ is non-vanishing, instead of
a condition on a Lagrange multiplier, one would get a tertiary constraint $\Upsilon=0$.
It is likely that it would not commute with $\CCpm^I$ and therefore its stability condition would fix a Lagrange multiplier,
as did the stability of $\Psi$ in the previous situation.
Comparing the resulting canonical structure to the previous one, one finds that the number of constraints
has increased by one. Whereas the role of $\Psi$ is played now by $\Upsilon$, $\Psi$ itself becomes now first class
removing one degree of freedom. This is precisely the expected canonical structure of a partially massless theory
where the additional gauge symmetry is generated by the secondary constraint $\Psi$!

\item
Even a simpler possibility is realized if the condition \eqref{stPsi} turns out to be satisfied automatically
on the constraint surface and upon taking into account all equations on the Lagrange multipliers.
This would imply that two second class constraints ($\Psi$ and one of the combinations of $\CCpm^I$)
converted into first class. This also removes one degree of freedom and therefore the resulting theory
can be considered as partially massless.
The difference with the previous case is that now one has {\it two} additional gauge symmetries and not just one.
It should also be clear that the condition that this happens is much stronger than the one leading to the previous case.

\item
Finally, it is possible that our theory represents an example of an irregular dynamical system \cite{Miskovic:2003ex}:
whereas for generic field configurations the stability condition \eqref{stPsi} fixes a Lagrange multiplier,
for a special class of solutions one of the last two possibilities is realized leading to additional gauge symmetries
and to the reduction of degrees of freedom.

\end{itemize}

\subsection{The main condition}

We provide some intermediate steps of the calculation of the stability condition \eqref{stPsi} in appendix
\ref{subap-stabpsi}.
The resulting equation takes the following form
\beq
&&\frac14\,{\eps_I}^{JK}\eps_{ab}\left\{E_0^I\Bigl[(\Lambda_+-\beta_+)\tppi{J}^a\tppi{K}^b
+2(\beta_+-\beta_-)\tppi{J}^a\tpmi{K}^b-(\Lambda_--\beta_-)\tpmi{J}^a\tpmi{K}^b\Bigr]
\right.
\nonumber\\
&&+\left.
2\Bigl[\( (\Lambda_+-\beta_+)\eepi{0}^I+(\beta_+-\beta_-)\eemi{0}^I\)\tppi{J}^a-
\( (\Lambda_- -\beta_-)\eemi{0}^I-(\beta_+-\beta_-)\eepi{0}^I\)\tpmi{J}^a\Bigr]\tP^b_K\right\}.
\nonumber\\
&&-
2{\eps^I}_{JK}\(\beta_+\tppi{I}^a-\beta_-\tpmi{I}^a\)\hom_a^J\hom_0^K
+ \eps_{IJK}\(\beta_+\eepi{0}^I-\beta_-\eemi{0}^I\)\eps^{ab}\hom_a^J\hom_b^K=0.
\label{eqstabPsi}
\eeq
It is this equation that we should analyze to determine the canonical structure of the theory and, in particular,
whether it can be partially massless.

For generic values of parameters, the first two lines in \eqref{eqstabPsi}
produce a linear combination of $\eepi{0}^I$ and $\eemi{0}^I$
with non-vanishing coefficients.
After substitution of the solution of \eqref{cone0} (see the next subsection),
which fixes a part of these Lagrange multipliers, it reduces to the expression \eqref{noconterms}.
Thus, as was initially assumed, the equation fixes
one of the Lagrange multipliers and we do not find anything new.

However, precisely for the special case \eqref{pmchoice}, all connection independent terms disappear and
the equation crucially simplifies, reducing to just two terms
\be
 \eps_{IJK}\(\eepi{0}^I-\eemi{0}^I\)\eps^{ab}\hom_a^J\hom_b^K-2{\eps^I}_{JK}\(\tppi{I}^a-\tpmi{I}^a\)\hom_a^J\hom_0^K=0.
\label{dtpsi}
\ee
Thus, we indeed find that the choice of parameters \eqref{pmchoice} is distinguished. But the found cancelations are not enough
yet to conclude that the theory is partially massless.
There are still two terms \eqref{dtpsi} which must be analyzed. To this end, we should take into account the conditions
on the Lagrange multipliers \eqref{cone0}, \eqref{conEom} and \eqref{stabS}.
In fact, it turns out to be possible to solve all these conditions explicitly, as we show in the next subsection.

\subsection{Solution for Lagrange multipliers}

To write down solution of the equations for Lagrange multipliers, we need to introduce some new notations.
First, we define the two metrics
\be
\mett_{ab}\equiv\eta_{IJ} \eepi{a}^I \eemi{b}^J,
\qquad
\metb_{ab}\equiv\eta_{IJ}E_a^I E_b^J,
\ee
which are both assumed to be invertible. Their non-vanishing
determinants will be denoted by $\mett$ and $\metb$, respectively.
Note that $\mett_{ab}$ is symmetric as a consequence of the constraint $\CS$ \eqref{defS}.
Next, we introduce the following variables
\be
\tXpm^I=\hf\,\eps^{IJK}\eps_{ab}\tppmi{J}^a\tppmi{K}^b,
\qquad
\tX^I=\hf\,\eps^{IJK}\eps_{ab}\tP^a_J\tP^b_K.
\label{defX}
\ee
They satisfy the orthogonality relations \eqref{orthX} which imply that $(\eepi{1}^I,\eepi{2}^I,\tXp^I)$, $(\eemi{1}^I,\eemi{2}^I,\tXm^I)$
or $(E_1^I,E_2^I,\tX^I)$
can be used as a basis in the tangent vector space. Therefore, we can decompose the Lagrange multipliers $\eepmi{0}^I$
with respect to one of these bases, which allows to introduce the lapse and the shift variables,
\be
\eepmi{0}^I=\Npm\tXpm^I+\Npm^a\eepmi{a}^I.
\label{defNN}
\ee
It is also convenient to define their diagonal and off-diagonal combinations by
$\Npm^a=\hf\,(N^a\pm\hN^a)$ and $\Npm=\hf\,(N\pm\hN)$.
The virtue of the change of variables \eqref{defNN} is that, as we will see,
it allows to disentangle the diagonal shift $N^a$ from all stability conditions.

Now it is easy to solve the condition \eqref{cone0}. Plugging the decomposition \eqref{defNN} into this equation,
one finds that it fixes the off-diagonal shift as
\be
\hN^a=\mett^{ab}\(\Nm(\eepi{b}\tXm)-\Np(\eemi{b}\tXp)\),
\label{solhN}
\ee
where $\mett^{ab}$  is the inverse of the metric $\mett_{ab}$.

The solution of the conditions \eqref{conEom} and \eqref{stabS}, which completely fix $\hom_0^I$, is a bit more involved.
We present it in appendix \ref{subap-solom0} and the result can be written in the following form
\be
\hom_0^I= N^a \hom_a^I+\(\Np\cfQpi{J}{a,I}+\Nm\cfQmi{J}{a,I}\)\hom_a^J,
\label{resomzero}
\ee
where the explicit expressions for the matrices $\cfQpmi{J}{a,I}$, which depend only on the canonical variables $\tppmi{I}^a$,
can be found in \eqref{defcfQ}

\subsection{Analysis of the main condition}

Plugging the solutions for Lagrange multipliers obtained in the previous subsection into \eqref{dtpsi},
one obtains the following equation
\be
\begin{split}
&
\eps_{IJK}\[\Np\(\tXp^I-\hf\,E_c^I \mett^{cd} (\eemi{d}\tXp)\)
-\Nm\(\tXm^I-\hf\,E_c^I \mett^{cd} (\eepi{d}\tXm)\)\]\eps^{ab}\hom_a^J\hom_b^K
\\
&
 -2{\eps^I}_{JK}\(\tppi{I}^a-\tpmi{I}^a\)\hom_a^J
\(\Np\cfQpi{L}{b,K}+\Nm\cfQmi{L}{b,K}\)\hom_b^L=0.
\end{split}
\label{redstab}
\ee
Note that all terms proportional to the diagonal shift $N^a$ canceled, in agreement with the fact that diagonal diffeomorphisms
remain unbroken gauge transformations. As a result, the only Lagrange multipliers entering the equation are the two lapses $\Npm$
and the stability condition takes the schematic form
\be
\Np\CW_+ +\Nm\CW_-=0.
\ee
The question now is whether this linear combination can degenerate into \eqref{degstabcon} where $\CN$ is
non-vanishing due to one of our non-degeneracy conditions. There are three such conditions which have been imposed so far,
which ensure the invertibility of $\eepi{\mu}^I$, $\eemi{\mu}^I$ and $E_\mu^I$.
The first two imply that $\Npm$ are non-vanishing, whereas the latter gives
\be
\begin{split}
\Ns\equiv\tX_I E_0^I=&\,(\tX\tCXp)\Np+(\tX\tCXm)\Nm
\\
=&\, \mett^{-1}\(\beta_+\gp(\tX\tXm)\Np+\beta_-\gm(\tX\tXp)\Nm\)\ne 0.
\end{split}
\label{defNs}
\ee

Let us now consider a special class of solutions which are close to a diagonal one, i.e. $\eepi{a}^I-\eemi{a}^I\sim\epsilon\ll 1$.
Then the equation \eqref{redstab} in the leading approximation reads
\be
\hN\eps_{IJK}\tX^I\eps^{ab}\hom_a^J\hom_b^K= O\(\epsilon N\hom_a\hom_b\).
\ee
Substituting \eqref{defX} into the l.h.s. and using \eqref{defPsi}, it can be rewritten as
$\hN\(\hom_a \tP^b\)\(\hom_b\tP^a\)$. Thus, it is generically non-vanishing and near the diagonal backgrounds
the equation does approximately degenerate into \eqref{degstabcon} with $\CN=\hN$.
But this combination of the Lagrange multipliers does not coincide neither with $\Npm$ nor with $\Ns$ \eqref{defNs},
which in this approximation becomes $\Ns\approx -4 \Lambda^2\mett N$.
As a result, the equation does not generate a tertiary constraint, but rather fixes $\hN$.
And once this happens even for a certain class of solutions, this excludes the second and third scenarios
presented in section \ref{subsec-scenario}. Hence, neither of the scenarios leading to a partially massless theory
is realized in three dimensions.

\section{Special backgrounds}
\label{sec-backgr}

Although the above analysis shows that the ghost-free bimetric gravity cannot be partially massless at full non-linear level
independently of the choice of parameters, for the special choice \eqref{pmchoice} suggested in \cite{Hassan:2012gz,Hassan:2012rq}
there is a whole class of backgrounds where the canonical structure changes.
Indeed, the stability condition \eqref{dtpsi} (or \eqref{redstab}) turns out to be automatically satisfied for
solutions which have
\be
\hat\omega_a^I=0.
\label{homzero}
\ee
Therefore, on this surface in the phase space the third scenario of section \ref{subsec-scenario} is realized
and two new gauge symmetries are generated.
Of course, outside the surface the constraints generating these additional symmetries become non-commuting and
we return to the previous picture. Thus, the total phase space corresponds to an irregular dynamical system
so that it is the fourth scenario which turns out to be relevant in this situation.

It is very important that, according to our result \eqref{redstab}, $\p_t\Psi\sim O(\hom_a^2)$, i.e. the time variation of
the secondary constraint vanishes {\it quadratically} with $\hom_a^I$.
Therefore, the two additional gauge symmetries removing one of the degrees of freedom are
not broken by linear fluctuations around the backgrounds satisfying \eqref{homzero}, but disappear only at quadratic level.
This fact explains the previous findings of partial masslessness
for linear perturbations around proportional backgrounds \eqref{propbackmet} \cite{Hassan:2012gz,Hassan:2012rq}.
Indeed, in the vielbein formalism the proportional backgrounds satisfy
\be
\eemi{\mu}^I=c\,\eepi{\mu}^I
\label{propbacktet}
\ee
and lead to $\hom_a^I=0$
because the spin-connection found from the Cartan equations depends homogeneously on the vielbein.

Furthermore, let us display all first class constraints existing on the surface \eqref{homzero}.
Their full set is given by
\be
\begin{split}
\CG_I=&\, \CGpi{I}+\CGmi{I},
\\
\CC_a=&\, \eepi{a}^I\CCpi{I}+\eemi{a}^I\CCmi{I}+\hom_a^I\hCG_I,
\\
\CH_+=&\, (\tXp\CCp)+(\tXp\eemi{a})\mett^{ab}(\eemi{b}\CCm)+\cfQpi{J}{a,I}\hom_a^J\hCG_I,
\\
\CH_-=&\, (\tXm\CCm)+(\tXm\eepi{a})\mett^{ab}(\eepi{b}\CCp)+\cfQmi{J}{a,I}\hom_a^J\hCG_I,
\\
\Psi'=&\, (\tP^a\hom_a)
+\frac{\Lambda}{4}\,\eps^{ab}(\hom_a \Theta_b)+\p_a\Sigma^a,
\end{split}
\label{fclass}
\ee
where $\Theta_a^I$ and $\Sigma^a$ are linear combinations of second class constraints computed
in \eqref{defTheta} and \eqref{defSigma}.
The first four of the above constraints are obtained as the functions multiplying
the undetermined Lagrange multipliers $\omega_a^I$, $N^a$ and $\Npm$, respectively, in the total Hamiltonian \eqref{hamilt}
after plugging there \eqref{defNN}, \eqref{solhN} and \eqref{resomzero}.
On the other hand, the last constraint promotes the secondary constraint $\Psi$ to a function which commutes with all other constraints.
Its construction requires the computation of the Dirac matrix of commutators of the second class constraints and its inverse,
which is quite involved and therefore differed to appendix \ref{ap-first}.

The resulting expression for $\Psi'$ might seem to be complicated.
But it provides a very simple explanation of another observation of \cite{Hassan:2012gz,Hassan:2012rq} that
the proportional backgrounds admit a gauge symmetry which rescales the parameter $c$ in \eqref{propbacktet}.
Indeed, if we perform a gauge transformation on the background fields, which is generated by $\Psi'$
with a constant gauge parameter, then the last two terms in \eqref{fclass} do not contribute:
the corresponding contributions are proportional either to $\hom_a^I$ or $\p_a\epsilon$ which both vanish.
Thus, we remain with
\be
\begin{array}{rlrl}
\delta_\epsilon\tppi{I}^a=&\, -\epsilon\(\tppi{I}^a+\tpmi{I}^a\),
\qquad &
\delta_\epsilon\tpmi{I}^a=&\, \epsilon\(\tppi{I}^a+\tpmi{I}^a\),
\\
\delta_\epsilon\ompi{a}^I=&\, 0,
\qquad &
\delta_\epsilon\ommi{a}^I=&\,0.
\end{array}
\ee
The meaning of this transformation becomes clear if one further restricts to the proportional backgrounds
\eqref{propbacktet}. Then it just changes the proportionality factor $c$
in agreement with findings of \cite{Hassan:2012gz,Hassan:2012rq} that this factor is not determined by equations of motion.
Our results however provide an extension of this simple observation and allow to compute the action of all symmetry generators
on linear fluctuations with arbitrary transformation parameters.

We repeat however that all these conclusions are valid only in the linear approximation around \eqref{homzero},
and already at quadratic level the additional gauge symmetries are not present.

\section{Conclusions}
\label{sec-concl}

In this work we have shown that independently of the choice of parameters zwei-dreibein gravity
(or the three-dimensional ghost-free bigravity in the dreibein formulation)
cannot become partially massless at non-linear level.

This result does not contradict the claims of \cite{Hassan:2012gz,Hassan:2012rq}
which only obtained necessary conditions on the parameter space of the theories.
Furthermore, our analysis provided an explanation for these claims from the point of view of
the complete canonical formulation of non-linear theory --- the parameter choice singled out in
\cite{Hassan:2012gz,Hassan:2012rq} is indeed special since precisely for this choice in the linear approximation
around proportional backgrounds the canonical structure turns out to change and corresponds
to a partially massless theory with one less degree of freedom. This reduction of degrees of freedom happens
because two constraints, which are second class at non-linear level, are converted into first class
after linearization. One of these constraints generates exactly the same transformations
which were used in \cite{Hassan:2012gz,Hassan:2012rq} as the criterion for partial masslessness.
An interesting open question is the geometric meaning of the transformations generated by the second constraint.

A very similar phenomenon was observed recently in other models of three-dimensional massive gravity
such as ``new massive gravity" and its various truncations and generalizations \cite{Bergshoeff:2009hq,Bergshoeff:2009aq,Deser:2009hb}.
For certain values of parameters these models also exhibit the linearization instability around some
special backgrounds, which, as in our case, corresponds to the emergence of the additional gauge
symmetry leading to partial masslessness. Furthermore, the mechanism responsible for this instability,
which was made explicit in
\cite{Blagojevic:2010ir,Afshar:2011qw,Hohm:2012vh,Deser:2012ci}\footnote{We thank Daniel Grumiller for drawing our attention to these works.},
turns out to be the same as in zwei-dreibein gravity --- on the special backgrounds
the stability condition of one of the constraints is fulfilled identically
thereby changing the canonical nature of two constraints.
This suggests that partial masslessness might always be an effect of such ``constraint bifurcation"
and do not extend to non-linear level.

However, we did not address this problem in four-dimensional models so that, in principle,
our work does not close the door to partial masslessness in this case.
Moreover, there exist some results which suggest that there are crucial differences
between massive gravity in $D=4$ and in other dimensions \cite{Zinoviev:2006im,mikaperso}.
In particular, this appears to be the case in bigravity models, as follows from the analysis of \cite{Hassan:2013pca,Hassan:2014vja},
and therefore may have an impact of the issue of partial masslessness.
Nevertheless, the canonical structure of ghost-free bigravity in four dimensions is very similar to the one discussed in this work
and the issue again reduces to the nature of the stability condition for the secondary constraint.
This constraint was found explicitly in \cite{Alexandrov:2013rxa} and, like \eqref{defPsi}, is also proportional to
the off-diagonal spin-connection $\hom_a$. Taking this into account as well as the findings of
\cite{deRham:2013wv,Deser:2013gpa,Joung:2014aba},
one might expect that our results here should be generalizable to four dimensions as well.
This certainly deserves further investigations.

\section*{Acknowledgments}
This research has received
funding from the European Research Council under
the European Communitys Seventh Framework Programme
(FP7/2007-2013 Grant Agreement no. 307934, "NIRG" starting grant).
CD thanks the Laboratoire Charles Coulomb of Montpellier for hospitality during a visit
leading to this work. SA thanks the Institut d'Astrophysique de Paris and the IH\'ES for the same.
We thank M. von Strauss for discussions.

\appendix

\section{Conventions}
\label{ap-conv}

In this paper we use the following conventions.
Components of spacetime tensors are labeled by greek indices $\mu,\nu,\dots=0,1,2$.
Their spatial components are labeled by latin indices
from the beginning of the alphabet $a,b,\dots =1,2$.
Components of the tangent space tensors are labeled by capital latin indices $I,J,\dots =0,1,2$,
which are raised and lowered by the flat Minkowski metric $\eta_{IJ}=\diag(-,+,+)$.
The Levi-Civita symbol with flat indices is normalized as $\eps_{012}=1$,
and contraction of two such symbols gives $\eps^{IJK}\eps_{KLM}=-\delta^I_L\delta^J_M+\delta^I_M\delta^J_L$.
On the other hand, for the antisymmetric tensor density with spacetime indices
we set $\eps^{0ab}=\eps^{ab}$. The symmetrization and anti-symmetrization
of indices denoted by $\{\cdot\, \cdot\}$ and $[\cdot\,\cdot]$, respectively, are both taken with weight 1/2.
The contraction of two tangent space vectors $\eta_{IJ}A^IB^J$ is sometimes denoted by $(AB)$.

Starting from the connection one-form $\omega^{IJ}$ we define
\be
\omega^I\equiv \hf\, {\eps^I}_{JK}\omega^{JK}.
\ee
As a result, its curvature
\be
F^{IJ}=\de\omega^{IJ}+{\omega^I}_K\wedge\omega^{KJ}
\ee
gives
\be
F^I\equiv \hf\, {\eps^I}_{JK}F^{JK}=\de\omega^I+\hf\,{\eps^I}_{KL}\omega^K\wedge \omega^L.
\ee
We also define its antisymmetric trace
\be
F\equiv \hf\, \eps^{ab}F_{ab}^I=\eps^{ab}\p_a\omega_b^I+\hf\, \eps^{ab}{\eps^I}_{KL}\omega_a^K \omega_b^L
\ee
and covariant derivative
\be
D_\mu A^I =\p_\mu A^I+{\eps^I}_{JK}\omega_\mu^J A^K.
\ee

\section{Constraint algebra}
\label{ap-algebra}

For any constraint $\Phi_\alpha$, where $\alpha$ is any set of indices, let us define its smeared version as follows
\be
\Phi(f)=\int \de \xc^2 f^\alpha \Phi_\alpha
\ee
with $f^\alpha$ being a set of arbitrary functions. Then, using this convention,
the algebra of the smeared primary constraints is given by
\begin{subequations}
\beq
\{\CGp(n),\CGm(m)\}&=& 0,
\\
\{\CGp(n),\CGp(m)\}&=& \CGp(n\times m),
\qquad
\{\CGm(n),\CGm(m)\}= \CGm(n\times m),
\\
\{\CGp(n),\CCp(N)\}&=& \CCp(n\times N)
\\
&&
+\int \de^2\xc\, \eps_{ab}\Bigl( \beta_+ \( (nN)(\tpp^a\tpm^b)-(n\tpp^a)(N\tpm^b)\) -\beta_- (n\tpm^a)(N\tpm^b)\Bigr),
\nonumber\\
\{\CGm(n),\CCp(N)\}&=& -\int \de^2\xc\, \eps_{ab}\Bigl( \beta_+ \( (nN)(\tpp^a\tpm^b)-(n\tpp^a)(N\tpm^b)\) -\beta_- (n\tpm^a)(N\tpm^b)\Bigr),
\\
\{\CGm(n),\CCm(N)\}&=& \CCmi{I}(n\times N)
\\
&&
-\int \de^2\xc\, \eps_{ab}\Bigl( \beta_- \( (nN)(\tpp^a\tpm^b)-(N\tpp^a)(n\tpm^b)\) -\beta_+ (N\tpp^a)(n\tpp^b)\Bigr),
\nonumber\\
\{\CGp(n),\CCm(N)\}&=&
\int \de^2\xc\, \eps_{ab}\Bigl( \beta_- \( (nN)(\tpp^a\tpm^b)-(N\tpp^a)(n\tpm^b)\) -\beta_+ (N\tpp^a)(n\tpp^b)\Bigr),
\\
\{\CCp(N),\CCp(M)\}&=& -\Lambda_+\CGp(N\times M)+\beta_+\CGm(N\times M)
\\
&&
+2\beta_+\int\de^2\xc\Bigl((N\hom_a)(M\tpm^a)-(M\hom_a)(N\tpm^a)\Bigr),
\nonumber\\
\{\CCm(N),\CCm(M)\}&=& \beta_-\CGp(N\times M)-\Lambda_-\CGm(N\times M)
\\
&&
-2\beta_-\int\de^2\xc\Bigl((N\hom_a)(M\tpp^a)-(M\hom_a)(N\tpp^a)\Bigr),
\nonumber\\
\{\CCp(N),\CCm(M)\}&=& 2\int\de^2\xc\Bigl(\beta_+\((NM)(\tpp^a\hom_a)-(N\hom_a)(M\tpp^a)\)
\\
&&
+\beta_-\((NM)(\tpm^a\hom_a)-(M\hom_a)(N\tpm^a)\)\Bigr),
\nonumber
\eeq
\label{algfirst}
\end{subequations}
where
\be
(n\times m)^I={\eps^I}_{JK}n^J m^K.
\ee
The secondary constraints produce the following commutators
\begin{subequations}
\beq
\{\CGp(n),\CS\}&=& -\eps_{ab}{\eps_I}^{JK}n^I\tppi{J}^a\tpmi{K}^b,
\\
\{\CGm(n),\CS\}&=& \eps_{ab}{\eps_I}^{JK}n^I\tppi{J}^a\tpmi{K}^b,
\\
\{\CCp(N),\CS\}&=& -\tpmi{I}^a\Dp_a N^I,
\\
\{\CCm(N),\CS\}&=& \tppi{I}^a\Dm_a N^I,
\\
\{\CGp(n),\Psi\}&=& \hf\(\beta_+\tppi{I}^a \Dm_a n^I+\beta_-\tpmi{I}^a \Dp_a n^I\),
\\
\{\CGm(n),\Psi\}&=& -\hf\(\beta_+\tppi{I}^a \Dm_a n^I+\beta_-\tpmi{I}^a \Dp_a n^I\),
\\
\{\CCp(N),\Psi\}&=& \beta_+\eps^{ab}\hom_a^I\Dp_b N_I
+\hf\,\eps^{IJK}N_I\eps_{ab}\((\Lambda_+-\beta_+)\tppi{J}^a+(\beta_+-\beta_-)\tpmi{J}^a\)\tP^b_K,
\\
\{\CCm(N),\Psi\}&=& \beta_-\eps^{ab}\hom_a^I\Dm_b N_I
-\hf\,\eps^{IJK}N_I\eps_{ab}\((\Lambda_- -\beta_-)\tpmi{J}^a-(\beta_+-\beta_-)\tppi{J}^a\)\tP^b_K,
\\
\{\CS(\kappa),\Psi\}&=& -\frac{\kappa}{2}\,(\beta_+-\beta_-)\CS.
\eeq
\label{algsecond}
\end{subequations}

\section{Details on calculations}
\label{ap-details}

\subsection{Useful identities}
\label{subap-ident}

\be
(\tXpm \tppm^a)=(\tX \tP^a)=0,
\label{orthX}
\ee
\be
(\tXpm\tXpm)=-\gpm,
\qquad
(\tXp\tXm)=-\mett,
\qquad
(\tX\tX)=-\metb,
\ee
\be
\tX^I=\beta_+^2\tXp^I+\beta_-^2\tXm^I+\beta_+\beta_-\eps^{IJK}\eps_{ab}\tppi{J}^a\tpmi{K}^b,
\ee
\be
\metb^{ab}E_a^I E_b^J=\eta^{IJ}+\metb^{-1}\tX^I\tX^J,
\label{decompident}
\ee
\be
\mett^{ab}\eepi{a}^I \eemi{b}^J=\eta^{IJ}+\mett^{-1}\tXm^I\tXp^J.
\ee

\subsection{Solution for $\hom_0^I$}
\label{subap-solom0}

Using the orthogonality relations \eqref{orthX}, it is straightforward to check that
the conditions \eqref{conEom} and \eqref{stabS} are solved by
\be
\hom_0^I=(E_0\hom_a)\metb^{ab} E_b^I+
\dth^{-1} \Omega\tX^I,
\label{solom}
\ee
where
\be
\begin{split}
\Omega=&\, {\eps^{I}}_{JK}\(\tpmi{I}^a\eepi{0}^J+\tppi{I}^a\eemi{0}^J\)\hom_a^K
+ (E_0\hom_a)\metb^{ab} \eps_{bc}\(\beta_+(\tpm^c\tXp)+\beta_-(\tpp^c\tXm)\),
\\
\dth=&\, -{\eps_I}^{JK}\eps_{ab}\tppi{J}^a\tpmi{K}^b\tX^I
\\
=&\, (\beta_+\beta_-)^{-1}\[\metb+\( (\beta_+^2\tXp+\beta_-^2\tXm)\tX\)\].
\end{split}
\ee
Furthermore, one can substitute \eqref{defNN} and \eqref{solhN} into \eqref{solom} and thus express the result in terms
of the lapse and shift variables. Using the properties listed
in the previous subsection and taking into account that
\be
(E_a\hom_b)-(E_b\hom_a)\sim\Psi\approx 0,
\ee
one arrives at the following expression
\be
\begin{split}
\hom_0^I=&\, N^a \hom_a^I+
\Bigl(\Np(\tCXp\hom_a) +\Nm(\tCXm\hom_a) \Bigr)\metb^{ab} E_b^I
+\dth^{-1}\hat\Omega \tX^I,
\end{split}
\label{resomzero-ap}
\ee
where
\beq
\tCXpm^I &=&\beta_\pm \tXpm^I \mp \hf\,(\tXpm\eempi{c})\mett^{cd}\(\beta_+\eepi{d}^I-\beta_-\eemi{d}^I\),
\\
\hat\Omega&=&  {\eps_I}^{JK}\hom_a^I\[ \Np\( \tpmi{J}^a\tXpi{K}+\tppi{J}^{\{a}\tpmi{K}^{b\}}\eps_{bc}\mett^{cd}(\eemi{d}\tXp) \)
\right.
\nonumber\\
&& \left.\qquad
+ \Nm\( \tppi{J}^a\tXmi{K}-\tppi{J}^{\{a}\tpmi{K}^{b\}}\eps_{bc}\mett^{cd}(\eepi{d}\tXm) \)
\]
\label{resDelt}\\
&&
-\Bigl(\Np (\tCXp\hom_a)+\Nm (\tCXm\hom_a) \Bigr)\metb^{ab}\Bigl(\beta_+(\tXp\eemi{b})+\beta_-(\tXm\eepi{b})\Bigr).
\nonumber
\eeq
This result has the same form as in \eqref{resomzero} with
\be
\begin{split}
\cfQpm^{\,a,IJ}=&\, \metb^{ab} E_b^I\tCXpm^{J}+\dth^{-1}\tX^I \[ \eps^{JKL}
\( \tpmpi{K}^a\tXpmi{L}\pm\tppi{K}^{\{a}\tpmi{L}^{b\}}\eps_{bc}\mett^{cd}(\eempi{d}\tXpm) \)
\right.\\
&\, \left.
-\tCXpm^{J} \metb^{ab}\Bigl(\beta_+(\tXp\eemi{b})+\beta_-(\tXm\eepi{b})\Bigr)\].
\end{split}
\label{defcfQ}
\ee

\subsection{Stability condition for $\Psi$}
\label{subap-stabpsi}

Using the commutation relations computed in \eqref{algsecond},
one obtains the following explicit form of the stability condition \eqref{stPsi}
\beq
&&
\hf\,{\eps_I}^{JK}\eps_{ab}\Bigl[\( (\Lambda_+-\beta_+)\eepi{0}^I+(\beta_+-\beta_-)\eemi{0}^I\)\tppi{J}^a-
\( (\Lambda_- -\beta_-)\eemi{0}^I-(\beta_+-\beta_-)\eepi{0}^I\)\tpmi{J}^a\Bigr]\tP^b_K
\nonumber\\
&&+
\(\beta_+\tppi{I}^a \Dm_a +\beta_-\tpmi{I}^a \Dp_a \)\hom_0^I
+\eta_{IJ}\eps^{ab}\hom_a^I\(\beta_+\Dp_b \eepi{0}^J+\beta_-\Dm_b \eemi{0}^J\)
=0.
\eeq
Moreover, it is possible to exclude all terms involving derivatives.
To this end, we rewrite the second line as
\be
\begin{split}
&
\eps^{ab}\p_a\( (E_b\hom_0)-(E_0\hom_b)  \)-\beta_+(\hom_0\CGp)-\beta_-(\hom_0\CGm)
+\hf\,E_0^I\( F_I(\omp)-F_I(\omm)\)
\\
&
-2{\eps^I}_{JK}\(\beta_+\tppi{I}^a-\beta_-\tpmi{I}^a\)\hom_a^J\hom_0^K
+ \eps_{IJK}\(\beta_+\eepi{0}^I-\beta_-\eemi{0}^I\)\eps^{ab}\hom_a^J\hom_b^K.
\end{split}
\ee
Taking into account \eqref{conEom}, the first term can be dropped. Then, using $\CCp^I-\CCm^I$,
one finally arrives at the equation given in \eqref{eqstabPsi}.
Furthermore, upon using \eqref{defNN} and \eqref{solhN}, the first two lines independent of the connection
can be rewritten as follows
\be
\begin{split}
& \(3\beta_+\beta_--\beta_-\Lambda_+-2\beta_+^2\)\(\Np \eps^{ac}\eps^{bd}\gpi{ab}\mett_{cd}+\Nm\mett\)+3\beta_+(\beta_+-\Lambda_+)\Np\gp
\\
-&\,\(3\beta_+\beta_--\beta_+\Lambda_--2\beta_-^2\)\(\Nm \eps^{ac}\eps^{bd}\gmi{ab}\mett_{cd}+\Np\mett\)-3\beta_-(\beta_- -\Lambda_-)\Nm\gm,
\end{split}
\label{noconterms}
\ee
which demonstrates that they do not vanish for generic values of the parameters.

\subsection{First class constraint for special backgrounds}
\label{ap-first}

In the degenerate case $\hom_a^I=0$ the list of second class constraints is given by $\Phi_\alpha=(\CS,\hCC_a,\hCG_I)$ where
\be
\begin{split}
\CS=&\, \eps_{ab}(\tpp^a\tpm^b),
\\
\hCC_a=&\, \eepi{a}^I \CCpi{I}- \eemi{a}^I \CCmi{I}= \eepi{a}^I F_I(\omp)-\eemi{a}^I F_I(\omm),
\\
\hCG_I=&\, \CGpi{I}-\CGmi{I}=\Dp_a \tppi{I}^a-\Dm_a \tpmi{I}^a.
\end{split}
\ee
Using the constraint algebra presented in appendix \ref{ap-algebra}, the matrix of their commutators is found to be
\be
\Delta_{\alpha\beta}(\xc,\yc)\equiv \{\Phi_\alpha(\xc),\Phi_\beta(\yc)\}
=\( \begin{array}{ccc}
0 &   \opB_a  &  -2\eps_{ab}{\eps_I}^{JK}\tppi{J}^a \tppi{K}^b
\\
-\opB_a^\dagger & 0 & -4\mett_{ab}\tP^b_I
\\
2\eps_{ab}{\eps_I}^{JK}\tppi{J}^a \tppi{K}^b & 4\mett_{ab}\tP^b_I & 0
\end{array}  \)\delta(\xc,\yc),
\ee
where $\opB_a$ is the differential operator
\be
\begin{split}
\opB_a=&\, 2\mett_{ab}\eps^{bc}(\yc)\p^{(\xc)}_c+2{\eps_I}^{JK}\omega_a^I \eps_{bc}\tppi{J}^b\tpmi{K}^c,
\\
\opB_a^\dagger=&\, 2\mett_{ab}\eps^{bc}(\xc)\p^{(\yc)}_c+2{\eps_I}^{JK}\omega_a^I \eps_{bc}\tppi{J}^b\tpmi{K}^c.
\end{split}
\ee
The inverse matrix reads as
\be
(\Delta^{-1})^{\alpha\beta}
=
\( \begin{array}{ccc}
0 &   0  &  -\frac{1}{4\metb}\, \tX^I
\\
0 & 0 & \frac{1}{4\metb}\,\mett^{ab}\metb_{bc}\tP^{c,I}
\\
\frac{1}{4\metb}\, \tX^I & -\frac{1}{4\metb}\,\mett^{ab}\metb_{bc}\tP^{c,I} & \opD^{IJ}
\end{array}  \)\delta(\xc,\yc),
\label{invDirac}
\ee
where $\opD^{IJ}$ is another operator
\be
\opD^{IJ}=\frU^{IJ,a}(\yc)\p_a^{(\xc)}-\frU^{IJ,a}(\xc)\p_a^{(\yc)}+\frV^{IJ}
\ee
with
\be
\frU^{IJ,a}=\frac{\metb^{ab}}{4\dth}\( E_b^{(I}\tX^{J)}-\dth^{-1}\(E_b(\tX_++\tX_-)\)\tX^I \tX^J\),
\ee
\beq
\frV^{IJ}&=&\frac{1}{8\dth}\,\eps^{IJK}\biggl[
\tP^a_K\p_a\log \dth-\p_a \tP^a_K-2\tP^a_K \metb^{cd}\(\eta_{LM} E_c^M-\dth^{-1}\(E_c(\tX_++\tX_-)\)\tX_L\)\p_d E_a^L
\biggr.
\nonumber
\\
&&
\biggl.
+2\tP^a_K\eps_{ab}\mett^{bc}\(\eps^{dg}\p_g \mett_{cd}+\omega_c^L{\eps_L}^{MN}\eps_{gf}\tppi{M}^g\tpmi{N}^f\)
\biggr].
\eeq

Although $\Psi$ commutes with all first class constraints, it does not commute with $\Phi_\alpha$.
Indeed, it is easy to find that
\beq
\{\Psi,\CS(\kappa)\}&=& 0,
\\
\{\Psi,\hCC(N)\}&=& -\Lambda\eta_{IJ}\eps^{ab}\hom_a^I\(\hat e_{c}^J\p_b N^c+N^c\( \Dp_b \eepi{c}^J-\Dm_b \eemi{c}^J\)\),
\\
\{\Psi,\hCG(n)\}&=& -\Lambda\(\tppi{I}^a \Dm_a n^I+\tpmi{I}^a \Dp_a n^I\).
\eeq
Nevertheless, it can be promoted to a first class constraint by constructing a linear combination with constraints $\Phi_\alpha$
as follows
\be
\Psi'=\Psi-\{\Psi,\Phi_\alpha\}\(\Delta^{-1}\)^{\alpha\beta}\Phi_\beta.
\ee
Using the result for the inverse Dirac matrix \eqref{invDirac}, one obtains
\beq
\Psi'
&=& (\tP^a\hom_a)
+\frac{\Lambda}{4}\,\eta_{IJ}\eps^{ab}\hom_a^I\biggl[\(\( \Dp_b \eepi{c}^J-\Dm_b \eemi{c}^J\)+\hat e_{c}^J\p_b\)
\(\mett^{cd}\eps_{dg} \metb^{gf}E_{f}^K\hCG_K\)
\biggr.
\nonumber\\
&&\biggr.\qquad
+2{\eps^J}_{KL}\hat e_b^K\(\metb^{-1} \tX^L\CS-\mett^{cd}\eps_{dg} \metb^{gf}E_{f}^L\hCC_b+4\opD^{LM}\hCG_M\)
\biggl]
\nonumber\\
&&
+ \p_a\[ -\frac{1}{4}\,\mett^{ab}\hCC_b+\tP^a_I\opD^{IJ}\hCG_J\].
\label{firstconstr}
\eeq
Using the property
\be
\begin{split}
\tP^a_I\opD^{IJ}=&\,\frac{1}{4\dth}\,\tX^J\mett^{ab} \(\eps^{cd}\p_d \mett_{bc}+\omega_b^L{\eps_L}^{MN}\eps_{cd}\tppi{M}^c\tpmi{N}^d\)
\\
&\,
+\frac{1}{4}\,\eps^{ab}\p_b\(\dth^{-1}\tX^J\)+\frac{1}{4\dth}\,\eps^{ab}\tX^J\p_b,
\end{split}
\ee
the last term reduces to
\be
\p_a\(\tP^a_I\opD^{IJ}\hCG_J\)=
\p_a\[\frac{\mett^{ab}}{4\dth} \(\eps^{cd}\p_d \mett_{bc}+\omega_b^L{\eps_L}^{MN}\eps_{cd}\tppi{M}^c\tpmi{N}^d\)\tX^I\hCG_I\].
\ee
Thus, the full first class constraint takes the following from
\be
\Psi'=(\tP^a\hom_a)
+\frac{\Lambda}{4}\,\eps^{ab}(\hom_a \Theta_b)+\p_a\Sigma^a,
\label{fclPsi}
\ee
where
\beq
\Theta_a^I&=&\(\( \Dp_a \eepi{c}^I-\Dm_a \eemi{c}^I\)+\hat e_{c}^I\p_a\)
\(\mett^{cd}\eps_{dg} \metb^{gf}E_{f}^I\hCG_I\)
\nonumber\\
&&
+2{\eps^I}_{JK}\hat e_a^J\(\metb^{-1} \tX^K\CS-\mett^{bc}\eps_{cd} \metb^{dg}E_{g}^K\hCC_b+4\(\frV^{KL}+\p_b\frU^{KL,b}+2\frU^{KL,b}\p_b\)\hCG_L\),
\label{defTheta}\\
\Sigma^a&=&\frac{\mett^{ab}}{4\dth} \(\eps^{cd}\p_d \mett_{bc}+\omega_b^L{\eps_L}^{MN}\eps_{cd}\tppi{M}^c\tpmi{N}^d\)\tX^I\hCG_I
-\frac{1}{4}\,\mett^{ab}\hCC_b.
\label{defSigma}
\eeq

\providecommand{\href}[2]{#2}\begingroup\raggedright\endgroup


\end{document}